# COMPARATIVE ANALYSIS OF MEL-FREQUENCY CEPSTRAL COEFFICIENTS AND WAVELET BASED AUDIO SIGNAL PROCESSING FOR EMOTION DETECTION AND MENTAL HEALTH ASSESSMENT IN SPOKEN SPEECH

Idoko Agbo[1], Dr Hoda El-Sayed[2] and Dr M.D Kamruzzan Sarker[3]

[1,2,3]Department of Computer Science, Bowie State University, Bowie MD United States

## ABSTRACT

*The intersection of technology and mental health has spurred innovative approaches to assessing emotional well-being, particularly through computational techniques applied to audio data analysis. This study explores the application of Convolutional Neural Network (CNN) and Long Short-Term Memory (LSTM) models on wavelet extracted features and Mel-frequency Cepstral Coefficients (MFCCs) for emotion detection from spoken speech. Data augmentation techniques, feature extraction, normalization, and model training were conducted to evaluate the models' performance in classifying emotional states. Results indicate that the CNN model achieved a higher accuracy of 61% compared to the LSTM model's accuracy of 56%. Both models demonstrated better performance in predicting specific emotions such as surprise and anger, leveraging distinct audio features like pitch and speed variations. Recommendations include further exploration of advanced data augmentation techniques, combined feature extraction methods, and the integration of linguistic analysis with speech characteristics for improved accuracy in mental health diagnostics. Collaboration for standardized dataset collection and sharing is recommended to foster advancements in affective computing and mental health care interventions.*

## 1. INTRODUCTION

In recent years, the intersection of technology and mental health has opened up new avenues for assessing and understanding emotional well-being, with a particular focus on leveraging computational techniques for analyzing spoken speech. This intersection has become a promising frontier, offering innovative solutions for uncovering valuable insights into individuals' emotional states and mental health conditions. Among the various computational methodologies, Mel-frequency Cepstral Coefficients (MFCCs) and Wavelet-based audio signal processing have emerged as prominent techniques for emotion detection and mental health assessment from spoken speech.

The importance of this topic stems from its profound implications for both research and practical applications in mental health care. Emotion detection from speech not only aids in understanding human behaviour but also facilitates the development of tools and interventions aimed at promoting mental well-being. Given the increasing prevalence of mental health disorders globally, there exists a pressing need for accurate and efficient methods of assessment and intervention. Computational speech analysis presents an objective and quantifiable means of assessing mental health, thus complementing traditional subjective evaluation methods and overcoming associated biases and inconsistencies.

The decision to embark on research in this area is motivated by several key factors. Firstly, traditional methods of mental health assessment often rely on self-reporting or subjective evaluations, which can be prone to bias and lack consistency. Computational approaches offer an objective and reliable alternative, enhancing the overall accuracy of mental health assessments. Secondly, advancements in computing technologies have made sophisticated audio signal processing techniques more accessible, enabling researchers to delve deeper into speech patterns and extract valuable information related to emotional states and mental health conditions.

Historically, the exploration of emotion detection from speech has evolved from early studies in psychology and linguistics, focusing on understanding the intricate relationship between language and emotions. Early approaches relied heavily on manual annotation and qualitative analysis of speech features. However, with the advent of computational methods, particularly in signal processing and machine learning, researchers have transitioned towards more systematic and quantitative approaches to emotion detection. The introduction of MFCCs and Wavelet transforms marked a significant advancement by providing efficient methods for representing and analyzing speech signals, ushering in a new era of research in emotion detection and mental health assessment.

In light of these developments, this research proposal aims to conduct a comparative analysis of MFCCs and Wavelet-based audio signal processing techniques for emotion detection and mental health assessment in spoken speech. Through systematic evaluation of the strengths and limitations of each approach, this study seeks to contribute to the growing body of knowledge in this field and provide insights that can inform the development of more effective tools and interventions for mental health care. Rigorous experimentation and analysis are envisioned to advance our understanding of how computational techniques can be leveraged to enhance mental health assessment and intervention strategies.

## 2. BACKGROUND

The "Emotion detection from text and speech" survey by Sailunaz, K., Dhaliwal, M., Rokne, J. et al. delves into the burgeoning field of emotion recognition, which holds immense potential for diverse applications ranging from human-computer interaction to robotic research. It explores the complexities of extracting emotions from various sources, such as speech and text, highlighting the challenges and methodologies researchers employ across different domains. Methodologically, the survey combines prosodic and spectral features for emotion detection from speech, employing classifiers like Gaussian Mixture Models (GMM), Support Vector Machines (SVM), Hidden Markov Models (HMM), and hybrid and ensemble systems. Significant findings reveal promising accuracies, with GMMs achieving 84% accuracy and SVMs reaching rates as high as 95.1% for specific databases. Despite these advancements, areas of expansion are identified, including the need for natural databases to enhance real-world applicability, exploring ranking orders for emotion-specific features, and delving into intensity, personality, and mood detection. The study's strengths lie in its inclusion of multiple databases and methodologies alongside the use of combined prosodic and spectral features. However, weaknesses stem from the extensive reliance on simulated databases, which may not accurately reflect real-world emotional expressions and intensities. Overall, the survey provides valuable insights into the current landscape of emotion detection research and outlines avenues for future exploration and refinement in this dynamic field.

The study on Emotion detection in speech using deep networks by M. R. Amer, B. Siddiquie, C. Richey and A. Divakaran introduces a novel staged hybrid model for detecting emotional content in speech. This model combines the strengths of discriminative and generative models to leverage both short-term temporal phenomena and long-range temporal dynamics. Methodologically,

Conditional Restrictive Boltzmann Machines (CRBM) and Conditional Random Fields (CRF) are employed as generative and discriminative models to learn and represent short-term and long-term temporal features. Evaluation is conducted on multiple audio-visual datasets, including AVEC, VAM, and SPD, showcasing the superiority of the proposed hybrid model over existing approaches. Significant findings reveal the efficacy of the CRBMs in learning feature representations from predefined and raw features, with the CRF-CRBM model performing best on the AVEC dataset due to its ability to capture long-term dynamics. However, deep learning methods improve feature representation in datasets like VAM and SPD, where long-term dynamics are less prominent, with simpler classifiers like SVM sufficient to capture emotions. The study suggests avenues for expansion, particularly in exploring emotion detection when linguistic content is absent from the speech. The strengths of the research lie in its proposition of a hybrid model that effectively captures temporal dynamics and its evaluation of a new dataset in noisy environments. However, limitations include a narrow set of classified emotions and emotion primitives. Overall, the study contributes to advancing emotion detection in speech by integrating deep network architectures and hybrid modelling techniques, paving the way for future research in this domain.

The study by C. Busso, S. Mariooryad, A. Metallinou and S. Narayanan titled "Iterative Feature Normalization Scheme for Automatic Emotion Detection from Speech" presents an iterative feature normalization (IFN) framework designed to enhance the robustness of emotion detection systems to speaker-dependent variations. The IFN approach aims to normalize features across speakers while preserving acoustic differences between emotional classes, thereby improving the accuracy of emotion detection from speech. Methodologically, the framework utilizes z-standardization as an affine transformation and linear kernel Support Vector Machines (SVM) for classification. Significant findings indicate that the IFN framework outperforms systems trained without normalization or with global normalization, achieving better accuracies in detecting emotional speech, especially in real-life, unconstrained recordings. The study suggests potential areas for expansion, including exploring other transformations such as feature warping and adapting the approach for multiparty interactions in noisy environments. Strengths of the study lie in its focus on identifying neutral speech with high precision, while weaknesses include limitations in recognizing multiple emotional labels. Overall, the research contributes to advancing emotion detection from a speech by introducing a practical normalization framework tailored to address speaker-dependent variations, with implications for practical applications in real-world settings.

The paper by Abdalla, M.I., Abobakr, H.M. and Gaafar, T.S, titled "DWT and MFCCs based Feature Extraction Methods for Isolated Word Recognition" introduces a novel method for feature extraction in speech recognition by combining Discrete Wavelet Transform (DWT) and Mel Frequency Cepstral Coefficients (MFCCs). The objective is to enhance recognition performance by leveraging additional features extracted from the signal. Methodologically, the speech signal is decomposed into frequency channels using wavelet transform, and then MFCCs are calculated for each channel. The resulting features are concatenated to generate a new set of features for classification. The study employs Neural Networks (NN) for classification and evaluates the proposed method using speech samples from 15 male speakers uttering ten isolated words each, representing the digits zero to nine. Major findings indicate a significant improvement in recognition rate, with the new method achieving a recognition rate of 99.6% compared to 99.2% using MFCCs alone. Potential areas for expansion include exploring methods to determine optimal neural network architectures and applying the approach to more extensive and diverse datasets beyond the limited set of English digits and speakers used in the study. The study's strengths lie in its innovative approach to feature extraction and its demonstrated improvement in recognition performance. However, limitations include using a small dataset consisting of isolated words and a limited number of speakers, which may affect the generalizability of the findings. Overall, the research contributes to advancing speech recognition techniques by introducing an effective feature extraction method that combines DWT and MFCCs, which has implications for improving accuracy in various speech recognition tasks.

The "Audio Textures: Theory and Applications" paper by Lie Lu, Liu Wenyin and Hong-Jiang Zhang, introduces a groundbreaking concept in audio media termed "audio textures," which enables the synthesis of long audio streams based on short example audio clips. Methodologically, the authors propose a three-step process for audio texture construction involving extracting basic patterns, creating variations, and concatenating variable patterns into a continuous sequence. Additionally, the study extends the concept to constrained audio texture synthesis for restoring missing parts in audio signals. Significant findings demonstrate the feasibility of constructing audio textures and their potential applications in background music, lullabies, game music, and audio restoration. The authors identify areas for expansion, including developing perceptual similarity measurements, controlling global perception, and incorporating additional features such as harmonics and tempo estimation. The strengths of the study lie in its innovative concept and encouraging results from subjective evaluations. However, limitations include focusing on a limited set of audio examples and the need for more robust similarity measurement techniques. The research contributes to advancing audio synthesis methods and opens avenues for further exploration in audio processing and related fields.

The "Wavelet Theory and Applications: A Literature Study" report by Merry, R.J.E., provides an in-depth overview of wavelet theory and its applications in signal analysis. Methodologically, the paper compares the wavelet transform to the Fourier transform, highlighting the advantages of wavelet analysis in capturing local frequency content while retaining time information. Significant findings indicate that wavelet analysis offers multiresolution capabilities, enabling a more detailed examination of signal characteristics compared to the Fourier transform. The continuous wavelet transform, calculated through convolution with a wavelet function, and the discrete wavelet transform, utilizing filter banks for signal analysis, are discussed in detail. The paper suggests potential areas for expansion, including numerical analysis, signal processing, control applications, and audio signal analysis. The study's strengths lie in its clear explanation of wavelet theory and its diverse applications, providing valuable insights for engineers and researchers. However, weaknesses include the need for more empirical data or case studies to validate the claims made. Overall, the research contributes to advancing the understanding of wavelet theory and its practical utility in various domains, highlighting its potential as a powerful tool in signal processing and analysis.

The paper by Tzanetakis, G., Essl, G. and Cook, P., titled "Audio Analysis using the Discrete Wavelet Transform", explores the application of the Discrete Wavelet Transform (DWT) in analyzing non-speech audio signals. Methodologically, the authors implement and evaluate two algorithms: one for automatic classification of various types of audio using the DWT and another for detecting beat attributes in music. Implemented within the MARSYAS framework, these algorithms showcase the potential of DWT in extracting valuable information from audio signals. Significant findings indicate promising results in automatic classification and beat detection tasks. The study suggests areas for expansion, including investigating other wavelet families and combining features from different analysis techniques to enhance classification accuracy. The strengths of the study lie in its utilization of DWT for feature extraction and classification alongside the reliable evaluation platform provided by the MARSYAS framework.

Additionally, comparing traditional feature extractors is a benchmark for assessing DWT-based approaches. However, limitations include reliance on synthetic and real-world stimuli for evaluation, which may only partially represent the diversity of non-speech audio signals. Future studies could address this by utilizing more extensive and more diverse datasets. Overall, the research contributes to advancing audio analysis techniques and underscores the potential of DWT in various applications.

The paper "A wavelet-based approach to emotion classification using EDA signals" by Huanghao Feng, Hosein M. Golshan, and Mohammad H. Mahoor presents an automated method for emotion classification in children based on electrodermal activity (EDA) signals. Methodologically, the authors employ continuous wavelet transforms (CWT) using the complex Morlet (C-Morlet) wavelet function on recorded EDA signals to conduct time-frequency analysis and generate a feature space for emotion recognition. The dataset consists of multimodal recordings of social and communicative behaviour from 100 children under 30 months old, annotated for three primary emotions: "Joy," "Boredom," and "Acceptance." The study employs a support vector machine (SVM) classifier for emotion classification and demonstrates improved performance when using wavelet-based features compared to other methods. Significant findings indicate successful recognition of the three emotions through manual annotation and quantitative analysis.

Additionally, the study evaluates different wavelet functions, confirming the superiority of the C-Morlet wavelet function. Areas for expansion include utilizing larger datasets for more robust evaluation and extending the study to include a broader range of emotions and participants. The study's strengths include the robust wavelet-based feature extraction method and the use of a powerful classifier. In contrast, weaknesses include the limited dataset size and reliance on manual annotation for emotion recognition. Overall, the research offers a promising approach to emotion classification using EDA signals, with potential applications in wearable assistive devices and intelligent human-computer interfaces.

Chang, K.H. authored a thesis titled "Speech Analysis Methodologies Towards Unobtrusive Mental Health Monitoring" The thesis explores speech analysis techniques for measuring various mental states, including affect, stress, and sleep deprivation, using non-invasive methods like mobile phones. Methodologically, the study integrates techniques from speech processing, psychology, human-computer interaction, and mobile computing systems. Initially, the author revisits emotion recognition methods by constructing an affective model with a naturalistic emotional speech dataset tailored for real-world applications. Then, leveraging speech production theory, the study investigates the influence of psychological states on glottal vibrational cycles, the source of speech production. Finally, the author develops the AMMON library, a low-footprint C library for widely available phones, facilitating richer human-computer interaction and healthcare technologies. The significant findings indicate successful emotion classification using speech analysis, with contributions spanning multiple domains. Areas for expansion include exploring other types of sensors and deep learning algorithms for enhanced emotion classification and investigating wearable devices for continuous emotion monitoring. The study's strengths include its interdisciplinary approach and practical implications for mental health monitoring. In contrast, its limitations include focusing on a specific set of emotions and excluding other emotional states or disorders. The research presents promising avenues for unobtrusive mental health monitoring using speech analysis methodologies.

## 3. METHODOLOGY

A comprehensive review of existing literature on emotion detection from speech, application of MFCCs and wavelet techniques in feature extraction from speech and approaches in emotion detection was conducted. The focus of the review was on identifying essential techniques, methodologies and algorithms employed in emotion detection and related datasets. This study leveraged the Crowd-sourced emotional multimodal actors ′dataset (Crema-D), Ryerson audio-visual database of emotional speech and song (Ravdess), Surrey audio-visual expressed emotion (Savee), and Toronto emotional speech set (Tess) as primary sources of data.

## 3.1. Data Visualization and Exploration

An initial exploratory data analysis was conducted on the collected dataset. This involved examining the data sources, assessing data quality and performing basic descriptive statistics and visualizations. Using a combination of 4 different datasets, a data frame for storing all emotions of the data frame is then used for feature extraction and model training.

To gain a comprehensive understanding of the audio data and facilitate insightful analysis, wave plots and spectrograms were generated as part of the data visualization and exploration process. Wave plots provide a visual representation of the amplitude of an audio signal over time, offering insights into the loudness variations at different time points within the audio. This visualization aids in identifying patterns related to speech characteristics and emotional expressions, which are crucial for subsequent analysis.

Spectrograms, on the other hand, offer a detailed visual representation of the spectrum of frequencies of the audio signal as they vary with time. By displaying the frequency content over time, spectrograms allow for the identification of specific frequency components present in the audio, including speech features and emotional cues. Together, these visualizations serve as valuable tools for initial data exploration, enabling researchers to identify relevant segments for further analysis and providing a visual context for subsequent computational techniques employed in emotion detection and mental health assessment.

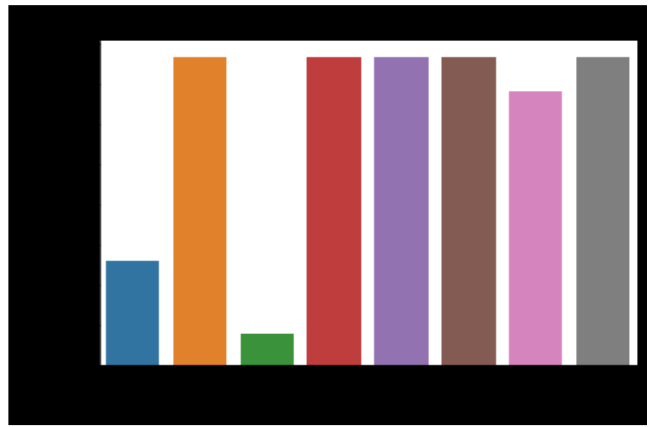

Fig 1: Plot the count of each emotions in our dataset

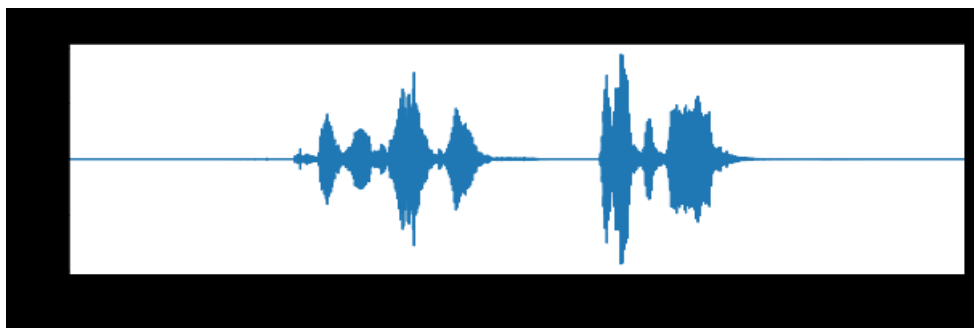

Fig 2: Waveplot for audio with fear emotion

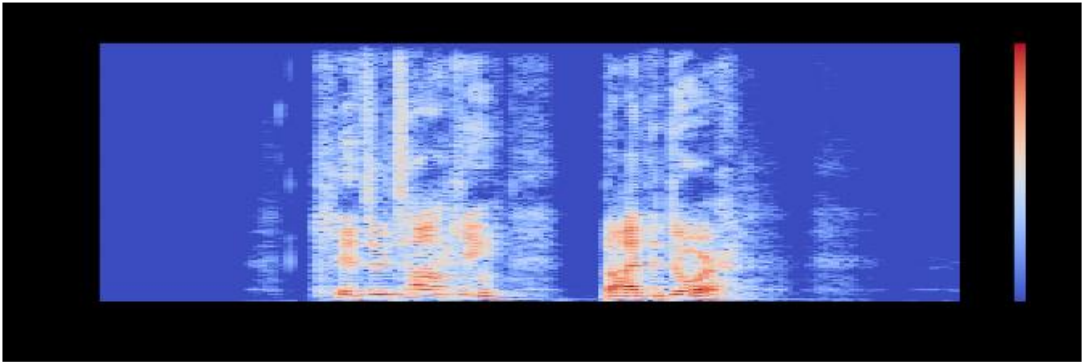

Fig 3: Spectrogram for audio with fear emotion

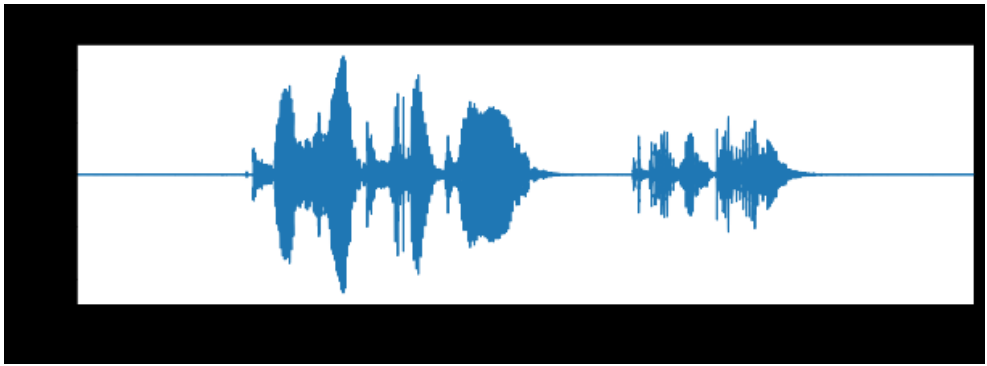

Fig 4: Waveplot for audio with sad emotion

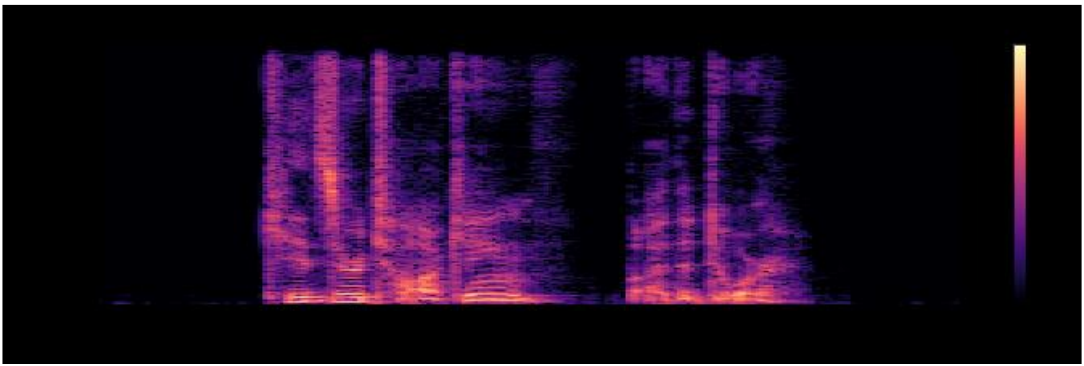

Fig 5: Spectrogram for audio with sad emotion

### 3.2. Data Augmentation

In the realm of audio data analysis, data augmentation plays a crucial role in enhancing the robustness and generalization capabilities of machine learning models. The process of data augmentation involves creating new synthetic data samples by introducing small perturbations to the initial training set, thereby expanding the diversity of the dataset. For our research, we have employed several key augmentation techniques, namely noise injection, stretching (changing speed), and pitch variations.

Noise injection involves adding controlled amounts of noise to the audio signal, simulating real-world environmental conditions and enhancing the model's ability to generalize to noisy environments. This technique is particularly effective in preventing overfitting and ensuring the model's performance across diverse audio inputs. By introducing variations in speed (stretching), we create synthetic data samples with altered temporal characteristics, helping the model learn invariant features across different speech rates and durations.

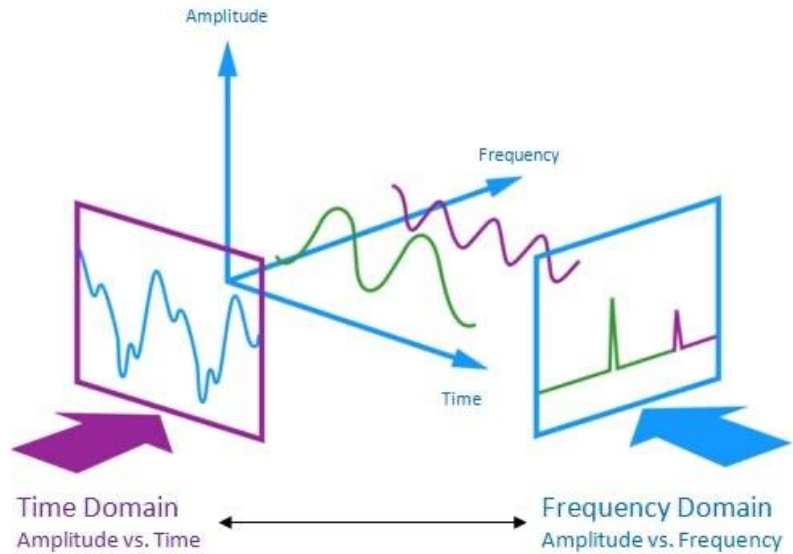

Fig 6: Audio signal as a 3 dimensional signal

Pitch variations, another essential augmentation technique, involve modifying the pitch of the audio signal while preserving its semantic content. This variation mimics natural fluctuations in pitch encountered in human speech, contributing to a more comprehensive understanding of speech patterns and emotional nuances across different pitch ranges.

The objective of employing these augmentation techniques is twofold: to create a more diverse and representative training dataset and to make our machine learning models invariant to these perturbations. It is crucial to ensure that the augmented samples retain the same labels as the original training samples, preserving the integrity of the dataset and aligning with our objective of training models for accurate emotion detection and mental health assessment.

While data augmentation techniques such as shifting, zooming, and rotating are common in image data, for audio data, techniques such as noise injection, stretching, and pitch variations are paramount in enhancing model generalization and mitigating overfitting issues.

### 3.3. Feature Extraction

Feature extraction plays a pivotal role in audio analysis by transforming raw audio data into a format that machine learning models can comprehend and analyze effectively. Unlike structured data formats readily interpretable by models, raw audio signals are complex and multidimensional, comprising temporal, amplitude, and frequency components. The audio signal is inherently a three-dimensional signal, where the axes correspond to time, amplitude, and frequency.

The temporal axis represents the progression of time, capturing the sequential nature of audio samples over time. Amplitude, represented by the vertical axis, denotes the intensity or loudness of the audio signal at different time points, providing crucial information about the audio's volume variations. Lastly, the frequency axis captures the pitch or tonal characteristics of the audio signal, representing the distribution of frequencies across the audio spectrum.

Given the intricate nature of audio signals, feature extraction techniques are employed to distill meaningful information and patterns from these multidimensional signals. Features extracted from audio signals may include spectral features such as Mel-frequency Cepstral Coefficients (MFCCs), which capture the spectral envelope of the audio signal and are instrumental in representing speech and emotional characteristics. Other features such as pitch, energy, and time-domain features like zero-crossing rate contribute to a comprehensive representation of audio characteristics relevant to emotion detection and mental health assessment.

By extracting relevant features, we transform the raw audio data into a structured and informative format that enables machine learning models to identify patterns, correlations, and emotional cues within the audio signals. This step is crucial for building accurate and robust models for emotion detection and mental health assessment, as it bridges the gap between raw audio data and model interpretability.

### 3.4. Modelling

After applying data augmentation techniques and extracting relevant features from each audio file, the data is now prepared for model training and testing. The first step involves normalization and splitting of the data into training and testing sets. Normalization is performed using sklearn's StandardScaler to scale the features, ensuring compatibility with the neural network model.

For the target labels (Y), one-hot encoding is applied as this is a multiclass classification problem, converting categorical labels into binary vectors for model training. The data is split using a random_state of 0 and shuffle set to True to ensure randomness in the data partitioning.
Following data preparation, a Convolutional Neural Network (CNN) model is constructed using Keras with a TensorFlow backend. The model architecture consists of several Conv1D layers followed by MaxPooling1D layers to extract relevant features and downsamples the data. Dropout layers are incorporated to prevent overfitting during training.

The model architecture includes:

- Conv1D layer with 256 filters, kernel size of 5, and ReLU activation
- MaxPooling1D layer with a pool size of 5 and strides of 2
- Additional Conv1D layers with decreasing filter sizes (256, 128, and 64) and similar configurations
- Dropout layers with dropout rates of 0.2 and 0.3 to reduce overfitting
- Flatten the layer to prepare the data for the dense layers
- Dense layers with ReLU activation to learn complex patterns
- Final Dense layer with a softmax activation for multiclass classification with 8 output classes corresponding to different emotional states

The model is compiled using the Adam optimizer and categorical cross-entropy loss function suitable for multiclass classification tasks. Model performance metrics such as accuracy are monitored during training and testing phases to evaluate the model's effectiveness in classifying emotional states from audio data.

## 4. RESULTS AND DISCUSSION

The application of Convolutional Neural Network (CNN) and Long Short-Term Memory (LSTM) models on wavelet extracted features and Mel-frequency Cepstral Coefficients (MFCCs) yielded varying accuracies, with the CNN model achieving a best-reported accuracy of 61% and the LSTM model reaching 56%. These accuracies reflect the models' abilities to classify emotional states from audio data based on distinct features extracted from the wavelet and MFCC representations. Notably, the models demonstrated better performance in predicting specific emotions such as surprise and anger, which aligns with the distinct audio characteristics associated with these emotions, such as variations in pitch and speed.

While the overall accuracy of 61% on the test data is considered decent, further improvements can be pursued through enhanced data augmentation techniques and exploration of additional feature extraction methods. Augmentation techniques such as varying noise levels, time stretching, and pitch modulation can introduce more diversity into the training data, potentially leading to better model generalization and improved accuracy across all emotional states.

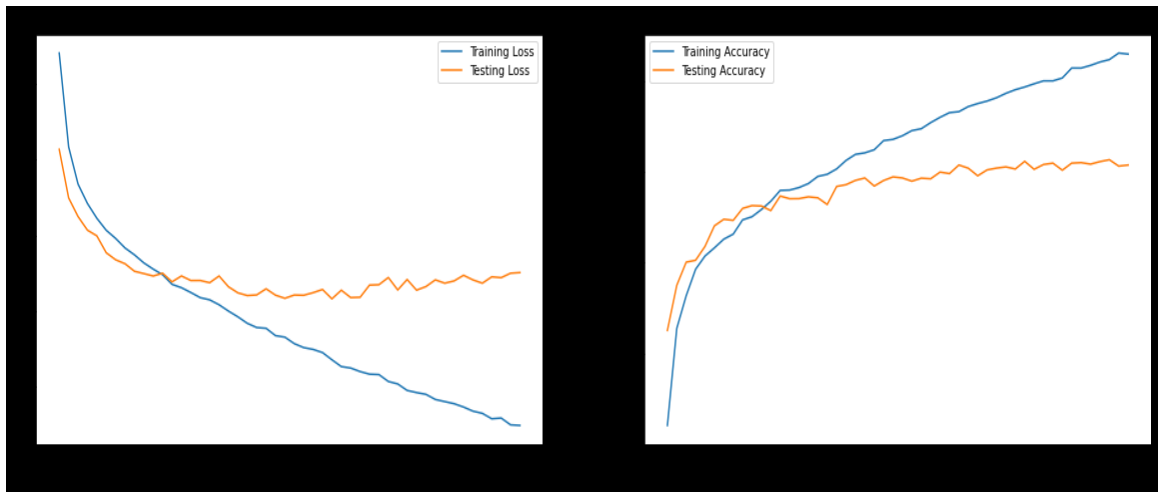

Fig 7: Accuracy of model on test data

The comparison with other machine learning models revealed mostly insignificant differences in performance, suggesting that the choice between CNN and LSTM models for this task may depend on factors such as computational resources and specific nuances of the dataset. It is worth noting that the computational cost increased significantly, especially with changes in epoch numbers ranging from 30 to 50 epochs, with minimal effects observed on overall model accuracy. This underscores the need for a balanced approach in model training duration and computational resource allocation, especially when marginal gains in accuracy are achieved with extended training epochs.

Moving forward, combining multiple feature extraction methods and exploring ensemble learning techniques could further enhance the model's performance, potentially pushing accuracy metrics beyond the current thresholds. Additionally, fine-tuning hyper-parameters and optimizing model architectures tailored to the nuances of emotional speech data could unlock additional gains in predictive accuracy, paving the way for more robust and reliable emotion detection systems based on audio analysis.

## 5. CONCLUSIONS AND RECOMMENDATIONS

In conclusion, our study explored the application of Convolutional Neural Network (CNN) and Long Short-Term Memory (LSTM) models on wavelet extracted features and Mel-frequency Cepstral Coefficients (MFCCs) for emotion detection from audio data. The CNN and LSTM models achieved best-reported accuracies of 61% and 56%, respectively, showcasing their ability to classify emotional states based on distinct audio features. While the models showed promise in predicting specific emotions like surprise and anger, further enhancements through advanced data augmentation techniques and exploration of combined feature extraction methods are recommended to improve overall accuracy. Additionally, comparisons with other machine learning models revealed marginal differences, emphasizing the need for tailored model architectures and computational resource allocation based on dataset nuances and training objectives. The computational cost of training, especially with extended epochs, highlights the importance of efficient model training strategies without compromising accuracy.

Moving forward, we recommend a holistic approach to emotion detection in audio data by combining speech content analysis with speech characteristics such as tone, pitch, and speed. Specifically, for the classification of anxiety and depression, future studies should consider integrating linguistic features such as words, choice of phrasing, and sentiment analysis with speech analysis techniques. Current medical methods for detecting anxiety and depression rely heavily on questionnaires and expert opinions, and a fusion of linguistic and speech analysis could significantly improve accuracy and reliability in mental health diagnostics.

Furthermore, we advocate for the collection of tailored datasets suited to emotion detection tasks, encompassing a diverse range of emotional states and speech contexts. Collaborative efforts towards standardizing data collection protocols and sharing benchmark datasets would facilitate advancements in the field of affective computing and support the development of robust mental health diagnostic tools and interventions. Additional studies focused on the provision of annotated datasets specifically designed for emotion classification tasks are crucial to drive innovation and progress in this vital area of research supporting mental health care.

## ACKNOWLEDGEMENTS

I would like to express sincere gratitude to all individuals and organizations who contributed to the completion of this research project. Appreciation goes to Dr Hoda El-Sayed and Dr. M.D Kamruzzan Sarker for their valuable guidance, feedback, and expertise throughout the research process.

**AUTHORS**

**Dr. Hoda El-Sayed** Dr. Hoda El-Sayed is a Professor in the Department of Computer Science at Bowie State University. Dr. El-Sayed's research interests include Parallel and Distributed Algorithms, High-Performance Computing and Programming. She received her D.Sc. in Computer Science from The George Washington University. Prior to joining Bowie State University, Dr. El-Sayed was a researcher in the High-Performance Computing Division at the National Institute of Standards and Technology (NIST).

**Dr. M.D Kamruzzaman Sarker** Dr. Sarker is an assistant professor in the Department of Computer Science at Bowie State University. Dr. Sarker's primary research theme is on Artificial Intelligence. His research interests include Trustworthiness of Deep learning models and space signal propagation optimization.

**Agbo Idoko** Agbo Idoko is a Doctoral Student in the Department of Computer Science at Bowie State University. Idoko's research interests include signal processing, machine learning and deep learning. He Received his bachelor's degree from Kwame Nkrumah University of science and Technology before joining Bowie State University.